\documentclass[aip,jcp,floats,amsmath,amssymb,preprint]{revtex4-1}

\usepackage{graphicx}
\usepackage{dcolumn}
\usepackage{bm}

\begin{document}

\preprint{(submitted to J. Chem. Phys.)}

\title{On the stability of "non-magic" endohedrally doped Si clusters:\\
A first-principles sampling study of $M$Si$_{16}^+$ ($M=$Ti,V,Cr)}

\author{Dennis Palagin}
\email{dennis.palagin@ch.tum.de}
\affiliation{Department Chemie, Technische Universit{\"a}t M{\"u}nchen, Lichtenbergstr. 4, D-85747 Garching, Germany}

\author{Matthias Gramzow}
\affiliation{Fritz-Haber-Institut der Max-Planck-Gesellschaft, Faradayweg 4-6, D-14195 Berlin, Germany}

\author{Karsten Reuter}
\affiliation{Department Chemie, Technische Universit{\"a}t M{\"u}nchen, Lichtenbergstr. 4, D-85747 Garching, Germany}
\affiliation{Fritz-Haber-Institut der Max-Planck-Gesellschaft, Faradayweg 4-6, D-14195 Berlin, Germany}

\date{April 5, 2011}

\begin{abstract}
Density-functional theory is used to study the geometric and electronic structure of cationic Si$_{16}^+$ clusters with a Ti, V or Cr dopant atom. Through unbiased global geometry optimization based on the basin-hopping approach we confirm that a Frank-Kasper polyhedron with the metal atom at the center represents the ground-state isomer for all three systems. The endohedral cage geometry is thus stabilized even though only VSi$_{16}^+$ achieves electronic shell closure within the prevalent spherical potential model. Our analysis of the electronic structure traces this diminished role of shell closure for the stabilization back to the adaptive capability of the metal-Si bonding, which is more the result of a complex hybridization than the orginally proposed mere formal charge transfer. The resulting flexibility of the metal-Si bond can help to stabilize also "non-magic" cage-dopant combinations, which suggests that a wider range of materials may eventually be cast into this useful geometry for cluster-assembled materials.
\end{abstract}

\maketitle

\section{Introduction}

Doping with endohedral metal atoms appears as a remarkable avenue to tailor the intrinsic properties of silicon clusters. \cite{jackson96,sun02,hiura01,kumar01,kumar02} In contrast to the compact geometries caused by the preferable $sp^3$ bonding in pure Si clusters \cite{jarrold91,ho98,jackson04,bai06}, the incorporation of even a single impurity atom can lead to the stabilization of otherwise unfavorable cage-like structures \cite{kumar06,reveles05,reveles06,torres07}. As in clathrates \cite{rachi05} or carbon nanostructures, these cages then represent appealing symmetric and unreactive building blocks for novel cluster-assembled materials with engineered properties. A prerequisite to a systematic synthesis of such materials are simple rules that rationalize which metal dopants stabilize cages and of which size. While this has been controversially discussed, one commonly agreed criterion for highly stable so-called "magic" clusters is geometric and electronic shell closure.\cite{jackson96,reveles05,reveles06} Here, the electronic manifold of a highly symmetric cage is viewed as states in a spherical potential, and particular stability is expected, if the electrons fill any one of the angular momentum shells, i.e. $1s$ ($2e^-$), $1p$ ($6e^-$), $1d$ ($10e^-$), $1f$ ($14e^-$), $2s$ ($2e^-$), $1g$ ($18e^-$), $2p$ ($6e^-$), $2d$ ($10e^-$) etc.\cite{jackson94} 

For a 16 Si atom endohedral Frank-Kasper (FK) polyhedron \cite{frank58,frank59} "magicity" would hence be predicted for a dopant atom donating 4 valence electrons, as the resulting $16 \times 4 + 4 = 68$ electrons achieve closure of the $2d$ shell. Within this model the known high stability of VSi$^+_{16}$ is thus naturally explained, if the nature of the bonding in the cluster is viewed in terms of a full formal charge transfer, i.e. "VSi$_{16}^+$ = V$^{5+}$ + Si$_{16}^{4-}$".\cite{torres07,koyasu05,koyasu07} Recently, however, Lau {\em et al.} deduced from their X-ray absorption spectroscopy data that also TiSi$^+_{16}$ and CrSi$^+_{16}$ with one valence electron less and more, respectively, stabilize in a cage geometry, with furthermore a highly similar local electronic structure around the dopant atom compared to the classic VSi$^+_{16}$ system.\cite{lau09} These findings motivate the present theoretical study, which follows a twofold goal. First, we perform a first-principles global geometry optimization of the three cluster systems $M$Si$_{16}^+$ ($M=$Ti,V,Cr) to firmly establish that the endohedral FK cage indeed represents the ground state geometry for all three dopant atoms. Second, we analyze the obtained electronic structure to obtain a more qualified view on the nature of the chemical bonding and elucidate the mechanism that stabilizes the cage despite the differing number of valence electrons in the three systems.

\section{Theory}

All calculations have been performed with the all-electron full-potential density-functional theory (DFT) code FHI-aims\cite{blum09}. Electronic exchange and correlation was treated within the generalized-gradient approximation functional due to Perdew, Burke and Ernzerhof (PBE)\cite{perdew96}. For comparison single-point calculations at the optimized PBE geometries were also performed on the hybrid functional level using the B3LYP \cite{vosko80} and PBE0 \cite{adamo99} functionals, without obtaining results that would lead to any conclusions different to the ones derived and presented below on the basis of the PBE data. FHI-aims employs basis sets consisting of atom-centered numerical orbitals. All sampling calculations are done with the "tier2" basis set, which contains 43 basis functions for Si, 67 basis functions for Ti, 88 basis functions for V, and 88 basis functions for Cr, respectively. The numerical integrations have been performed with the "tight" settings, which correspond to integration grids with 85, 97, 99, and 101 radial shells for Si, Ti, V, and Cr, respectively, in which the number of integration points is successively decreased from 434 for the outermost shell to 50 for the innermost one.\cite{blum09} For the ensuing electronic structure analysis of the optimized geometries the electron density was recomputed with an enlarged "tier3" basis set, which contains 64 basis functions for Si, 103 basis functions for Ti, 115 basis functions for V and 124 basis functions for Cr. Systematic convergence tests indicate that these settings are fully converged with respect to the target quantities (energetic difference of different isomers in the sampling runs; electron density distribution in the electronic structure analysis). This holds in particular for a central quantity of our analysis, the radial electron density distribution of the different doped cages. This quantity is defined as
\begin{equation}
n(r) \;=\; \int_{0}^{2\pi} \int_{0}^{\pi} r^2 n(r,\theta,\phi) \; {\rm sin}\theta \; d\theta d\phi \quad ,
\end{equation}
where $n(r, \theta, \phi)$ is the electron density at a given point at spherical coordinates $(r, \theta, \phi)$ away from the central dopant atom at $r=0$. To build a radial distribution of the electron density we calculate the surface integral, eq. (1), for a set of spheres of increasing radii, and then plot the obtained values as a function of the sphere radius. The numerical integration is hereby performed using a cubic $(400 \times 400 \times 400)$ volumetric data grid with 0.02\,{\AA} voxel width. The chosen finite integration radius and angle steps equal 0.02\,Bohr and ${\pi}$/$360$, respectively.

Local structure optimization is done using the Broyden-Fletcher-Goldfarb-Shanno method \cite{press07} relaxing all force components
to smaller than $10^{-2}$\,eV/{\AA}. To make sure that the cage-like geometry indeed represents the ground-state structure for all three dopant atoms we relied on basin-hopping (BH) based global geometry optimization \cite{wales97,wales00,gehrke09}. Within the BH idea the configuration space is explored by performing consecutive jumps from one local minimum of the potential energy surface (PES) to another. To achieve this, positions of atoms in the cluster are randomly perturbed in a so-called trial move followed by a local geometry optimization which brings the system again into a local PES minimum. A Metropolis-type acceptance rule is used to either accept or reject the jump into the PES minimum reached by the trial move. As specific BH implementation we chose collective as well as single-particle trial moves, in which all atoms (collective move) or a randomly picked atom (single-particle move) is displaced in a random direction. Two different starting points were used for all optimization runs: 1) All atoms are randomly positioned inside a box of dimension $(9 \times 9 \times 9)$\,{\AA}$^3$, or 2) the solution for the Thomson-problem \cite{thomson04} (how to put point charges on a sphere with minimal energy) is employed to position the Si atoms and then the doping metal atom is added at the center. Typical BH runs comprised of the order of 100 accepted trial moves, and unanimously identified the cage geometry as lowest energy structure regardless of the specific settings employed for the Metropolis rule or the single-particle/collective moves.

\section{Results}

\subsection{Cage-like ground state geometry}

\begin{figure}
\includegraphics[width=\linewidth]{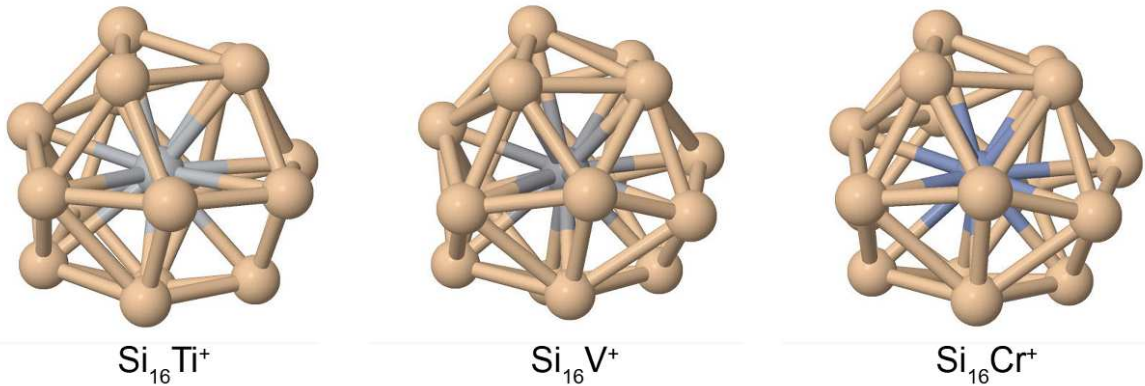}
\caption{Ball-and-stick views of the identified ground-state FK cage geometries.
\label{fig1}}
\end{figure}

In contrast to our preceding work on Si$_{20}$ fullerenes \cite{willand10}, our extended unbiased configuration searches confirm that the endohedral FK cage indeed represents the ground-state isomer for $M$Si$_{16}^+$ with all three dopant atoms, cf. Fig. \ref{fig1}. The "non-magicity" in case of Ti and Cr doping only expresses itself in form of a much reduced energetic gap to the second lowest energy isomer identified in the BH runs: For VSi$_{16}^+$ this gap amounts to 1.00\,eV, whereas for TiSi$_{16}^+$ and CrSi$_{16}^+$ it is only 0.01\,eV and 0.08\,eV, respectively. Within a 1.00\,eV range above the identified ground state we correspondingly found about 15 inequivalent isomers for the latter two systems. In the Cr-doped case all of them are capped CrSi$_{15}^+$ cages, for Ti more compact TiSi$_{16}^+$ cages are found within 0.1\,eV above the ground state. Above this mostly capped TiSi$_{15}^+$ structures are identified.

The incomplete shell-closure of TiSi$_{16}^+$ and CrSi$_{16}^+$ also shows up in the symmetry of the FK cage. Whereas the "ideal" VSi$_{16}^+$ cluster exhibits perfect $T_d$ symmetry, the cages with Ti and Cr dopants only exhibit $C_1$ symmetry. The distortions away from perfect $T_d$ symmetry are, however, only minor, as can best be seen from the $M$-Si bond distances within the cage. In a perfect FK polyhedron these distances fall into two closely spaced shells: One with four Si atoms that form a perfect tetrahedron, and slightly beyond that another one with 12 Si atoms that are all equidistant from the encapsulated metal atom. In the VSi$_{16}^+$ cluster, these two shells are located at distances of 2.54\,{\AA} and 2.81\,{\AA}, respectively. In the less symmetric TiSi$_{16}^+$ and CrSi$_{16}^+$ geometries the distortions lift the degeneracies of the two shells and we instead find $M$-Si distances spread over a range of 2.64\,{\AA} to 2.86\,{\AA} for Ti and over a range of 2.50\,{\AA} to 3.35\,{\AA} for Cr, respectively. Overall this leads in case of Ti doping to a slightly increased average cage radius of 2.78\,{\AA}, compared to the average $M$-Si distance of 2.74\,{\AA} in both VSi$_{16}^+$ and CrSi$_{16}^+$. 

Overall, the geometric differences between the three cages are thus rather small. We furthermore verified that these differences have only insignificant effects with respect to the discussion of the electronic structure of the cage presented in the following. Our analysis is therefore for all three doped cages, as well as the empty Si cage based on the symmetric geometry obtained for VSi$_{16}^+$, i.e. in all cases the geometry was kept as in VSi$_{16}^+$ and only the electron density was each time self-consistently computed. This procedure facilitates the qualitative discussion of the nature of the bonding and of the concomitant charge redistribution as it allows to directly subtract the electron densities obtained for the different dopants and, because of the higher symmetry, makes the presentation of radial electron distributions averaged over the solid angle more meaningful.

\subsection{Spherical potential model}

\begin{figure}
\includegraphics[width=\linewidth]{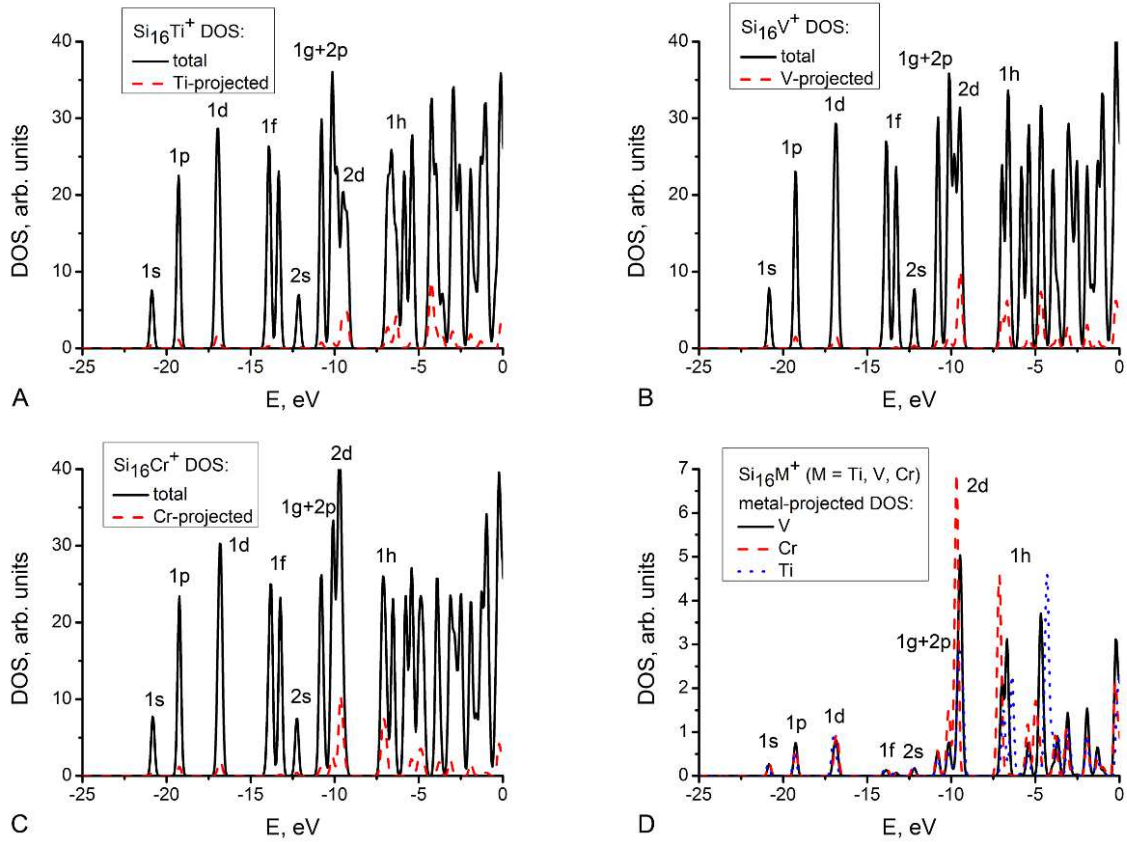}
\caption{Total density of states (DOS) and DOS projected on the metal dopant for A) TiSi$_{16}^+$, B) VSi$_{16}^+$, and C) CrSi$_{16}^+$. Panel D) directly compares the metal-projected DOS for the three cases to illustrate the varying degree of metal-Si hybridization. The zero-reference for the energy scale is the vacuum level, and the labels given to the different groups of states follow the notation of the spherical potential model (see text).
\label{fig2}}
\end{figure}

The prevalent model to rationalize the stability of doped Si cage geometries is the spherical potential model \cite{jackson96,jackson94,kumar06,reveles05,reveles06,torres07}, which has been discussed in detail for the "magic" VSi$_{16}^+$ cluster by Torres, Fernandez and Balbas \cite{torres07}. As a first step in our attempt to qualify the chemical bonding and stability in the "non-magic" FK clusters doped with Ti and Cr we first briefly recapitulate the essentials of this discussion. The spherical potential model exploits the near sphericity of the ideal FK polyhedron, which suggests to classify the electronic states in shells of a determined radial and angular momentum quantum number. The computed density of states (DOS) of VSi$_{16}^+$ shown in Fig. \ref{fig2}B demonstrates that the Kohn-Sham states indeed group into the expected sequence ($1s, 1p, 1d, 1f, 2s, 1g, 2p, 2d, 1h,\ldots$), with the 68 valence electrons exactly achieving closure of the $2d$ shell. Also more subtle features like the splitting into the different tetrahedral ($T_d$) sub-groups are perfectly obeyed, i.e. the different shells are sub-divided as $s (a_1), p (t_2), d (e + t_2), f (a_2 + t_1 + t_2), g (a_1 + e + t_1 + t_2), h (e + t_1 + 2t_2)$. Bonding to the transition metal dopant is predominantly expected via the $\pi$-type orbitals with one radial node ($2s, 2p, 2d$), with hybridization following an approximate $l$-selection rule, i.e. the dopant $3d$ and $4s$ valence orbitals mix with Si cage $d$ and $s$ $\pi$-orbitals, respectively. The metal-projected DOS contained in Fig. \ref{fig2}B proves that also this feature of the spherical potential model is fully reproduced by the actual computation. 

However, these features are not a specificity of the "magic" VSi$_{16}^+$ cluster, but instead inherent properties of the near-spherical FK polyhedral shape. As apparent from Fig. \ref{fig2}A and C essentially the same groupings of the Kohn-Sham states are equally obtained for the other two dopants, i.e. also here the electronic manifold is well described within the spherical potential model. Exactly as expected from the differing number of valence electrons, the only difference is that electronic shell closure is not achieved. In TiSi$_{16}^+$ (with 67 valence electrons) the highest energy state of the $2d$ shell is unoccupied, and in CrSi$_{16}^+$ (with 69 valence electrons) the lowest energy state of the $1h$ shell is occupied. This lifts many of the degeneracies within the different electronic shells, but the overall structure in terms of angular momentum shells is still preserved. Furthermore, as confirmed by our first-principles sampling calculations the endohedral FK polyhedron still represents the lowest-energy isomer for the "non-magic" TiSi$_{16}^+$ and CrSi$_{16}^+$. Electronic shell closure might thus be a criterion for enhanced stability, as e.g. reflected by the abundance of VSi$_{16}^+$ in the experimental mass spectra. However, it is not a necessary condition to stabilize the endohedral cage geometry {\em per se}.

\subsection{Charge transfer vs. hybridization}

\begin{figure}
\includegraphics[width=\linewidth]{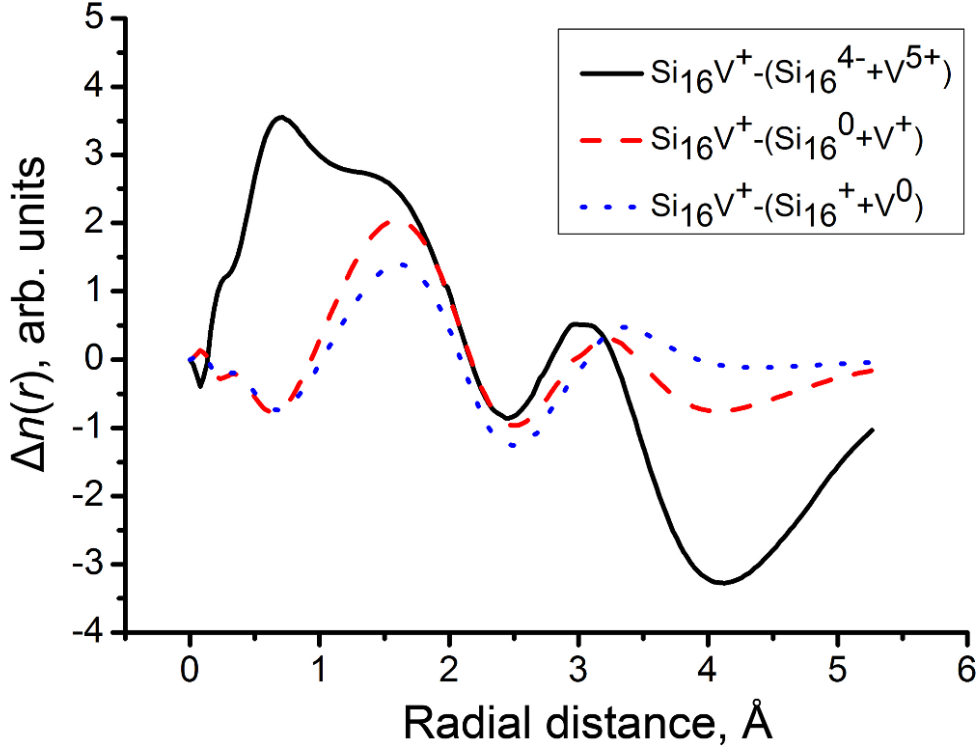}
\caption{Radial distribution, cf. eq. (1), of the electron density difference, $\Delta n(r) = n_{{\rm VSi}_{16}^+}(r) - n_{{\rm Si}_{16}^{4-}}(r) - n_{{\rm V}^{5+}}(r)$ (solid line), where $n_{{\rm VSi}_{16}^+}(r)$ is the electron density of the doped cage, $n_{{\rm Si}_{16}^{4-}}(r)$ the density of the empty Si cage, and $n_{{\rm V}^{5+}}(r)$ the density of the V cation. If the formal charge transfer picture was correct, $\Delta n(r)$ should be essentially zero throughout. Note the average cage radius, i.e. the position of the Si atoms, is at 2.74\,{\AA}. Additionally shown are other charge combinations of the two fragments (Si$_{16}$ + V$^+$, dashed line; Si$_{16}^+$ + V, dotted line).
\label{fig3}}
\end{figure}

Insight into the weakened role of electronic shell closure can come from a more qualified discussion of the nature of the chemical bond within the doped clusters. The simplified picture connected with the "magicity" of VSi$_{16}^+$ assumes a formal charge transfer of all V valence electrons to the Si cage manifold. This "formal" view is readily checked by evaluating the difference of the actually computed electron density of VSi$_{16}^+$ with respect to a mere superposition of the electron densities of an empty Si$_{16}^{4-}$ cage and a V$^{5+}$ cation. If the formal charge transfer picture was correct, then this electron density difference should be zero throughout. Figure \ref{fig3} shows this difference in form of the radial electron density distribution, i.e. averaged over the solid angle, cf. eq. (1). The largely negative values exhibited at radii larger than the average cage radius of 2.74\,{\AA} indicate that a formally $4-$ charged Si cage would contain much more electron density at the outside as compared to the real VSi$_{16}^+$ system, while simultaneously there would be much less charge in the inside (positive regions in Fig. \ref{fig3}). However, this does not simply indicate that a smaller formal charge transfer from metal to cage takes place. As illustrated in Fig. \ref{fig3} also other superpositions of differently charged empty cages and cations do not represent the real electron density well. This holds in particular for the radial region between $\sim 1 - 2$\,{\AA}, i.e. exactly the bonding region between central metal atom and cage. The metal-Si bonding is thus rather the result of a more complex hybridization than mere formal charge transfer.

\begin{figure}
\includegraphics[width=\linewidth]{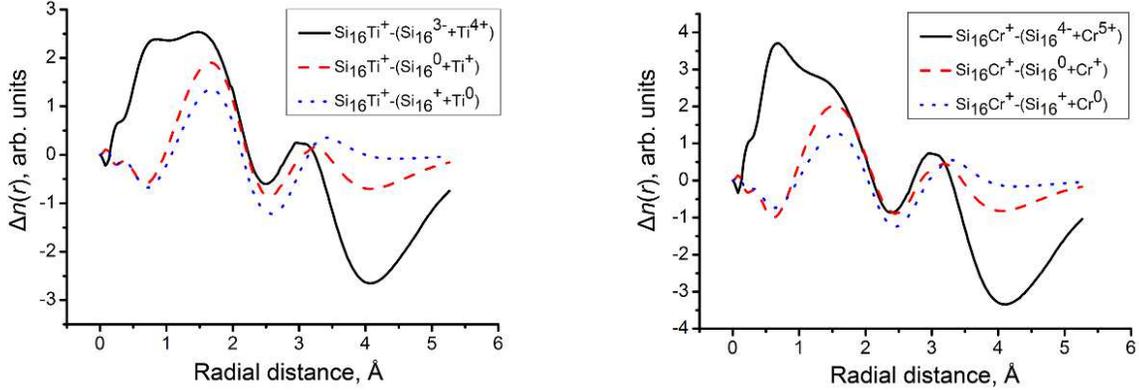}
\caption{Same as Fig. \ref{fig3}, but for TiSi$_{16}^+$ (left panel) and for CrSi$_{16}^+$ (right panel).
\label{fig4}}
\end{figure}

\begin{figure}
\includegraphics[width=\linewidth]{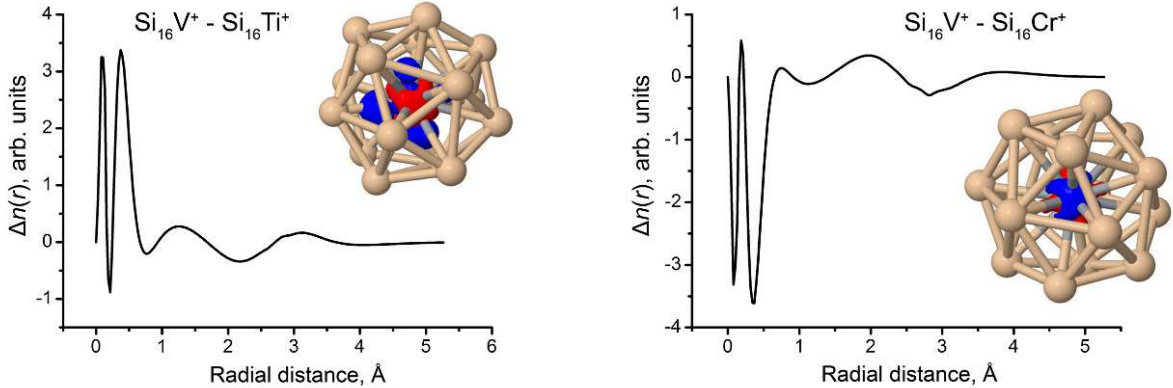}
\caption{Electron density difference VSi$_{16}^+$ - TiSi$_{16}^+$ (left panel) and VSi$_{16}^+$ - CrSi$_{16}^+$ (right panel). The radial electron density distribution, cf. eq. (1), as well as the 3D isosurface at $0.02$\,$e/${\AA}$^3$ in the inset demonstrate that the missing electron in the former and excess electron in the latter case are predominantly located around the central metal atom.
\label{fig5}}
\end{figure}

Equivalent results shown in Fig. \ref{fig4} are also obtained for the "non-magic" TiSi$_{16}^+$ and CrSi$_{16}^+$ clusters, which means that also there the real electron density of the endohedral cage cannot be fully rationalized in terms of a formal charge transfer. However, in all three dopant cases the true electron density outside the cage is best represented, i.e. the radial electron density difference distribution is closest to zero, for a charge combination of a positively charged Si cage and a neutral metal atom. This suggests that the different number of valence electrons in the three systems resides predominantly around the dopant. Figure \ref{fig5} demonstrates that this is indeed the case. Depicted is the electron density difference between VSi$_{16}^+$ and TiSi$_{16}^+$, as well as between VSi$_{16}^+$ and CrSi$_{16}^+$, which allows to locate the missing electron in TiSi$_{16}^+$ and the excess electron of CrSi$_{16}^+$ as compared to the "magic" VSi$_{16}^+$ cluster, respectively. In both cases this is close to the central metal atom.

A complementary view comes from the analysis of the projected DOS. For this Fig. \ref{fig3}D specifically compares the metal-projected DOS for the three doped cages. Interestingly, the metal contribution to the lower lying electronic shells up to the $2p$ shell is essentially the same in all cases. This is much different for the frontier shells $2d$ and $1h$, which are mostly responsible for the bonding between cage and dopant. Here, there is a clear trend of increasing metal weight to the states when going from TiSi$_{16}^+$ over VSi$_{16}^+$ to CrSi$_{16}^+$. If we add up these metal contributions over the occupied set of $2d$ and $1h$ states, we arrive at a total of 2.1 (Ti), 3.1 (V) and 3.7 (Cr) electrons in the three cases. Between Ti and V, as well as between V and Cr the metal dopant provides thus each time around one electron more to the hybridized states. The adapting degree of metal-Si hybridization hence compensates largely for the different total electron numbers. In other words, while from TiSi$_{16}^+$ over VSi$_{16}^+$ to CrSi$_{16}^+$ there is each time one more valence electron in the topmost $2d$ and $1h$ shells, the number of electrons that is actually assigned to the Si cage remains essentially the same. The cage therefore effectively "sees" similar charge numbers, as the adaptive ability of the orbitals that are predominantly responsible for the metal-silicon bonding can accommodate for the charge excess or deficit. Intriguingly, it is, however, not just one state, e.g. intuitively the one with the changed occupation, that is responsible for this. Instead it is the rehybridization of the entire set of $2d$ and $1h$ states, which effectively compensates for the "non-magicity". This adaptive capability diminishes the role of electronic shell closure and is in our view the main reason that helps to stabilize the endohedral cage geometry also for TiSi$_{16}^+$ and CrSi$_{16}^+$ despite their differing number of valence electrons.

\section{Conclusions}

In summary our DFT-based unbiased configuration searches confirm the preceding interpretation of Lau {\em et al.} \cite{lau09} that the Frank-Kasper polyhedron indeed represents the ground-state geometry for the series of doped TiSi$_{16}^+$, VSi$_{16}^+$ and CrSi$_{16}^+$ clusters. Endohedral doping can thus be used as avenue to stabilize cage-like Si$_{16}$ geometries. The electronic structure analysis demonstrates that all three systems are well described within the spherical potential model, i.e. the electronic manifold groups into states of defined radial and angular momentum number. Only the classic VSi$_{16}^+$ cluster achieves closure of the electronic $2d$ shell, while the varying number of valence electrons leads to an unoccupied $2d$ state in case of TiSi$_{16}^+$ and an occupied $1h$ state in case of CrSi$_{16}^+$. Shell closure is thus not a necessary condition for the stabilization of the cage-like geometry.

We attribute this diminished role of shell closure for the stabilization to the adaptive capability of the metal-Si bonding, which is more the result of a complex hybridization than the mere formal charge transfer picture originally proposed in connection with the spherical potential model. This adaptive capability allows to locate the deficient electron in case of TiSi$_{16}^+$, as well as the excess electron in case of CrSi$_{16}^+$ predominantly around the metal dopant. The effective charge assigned to the Si cage is then essentially the same in the three systems, i.e. the rehybridization of the $2d$ and $1h$ shells compensates for the "non-magicity". While electronic shell closure is still certainly a criterion for particularly enhanced stability, the flexibility of the metal-Si bond can help to stabilize also other cage-dopant combinations than predicted by this simple rule. This indicates the exciting prospect that a wider range of materials may eventually be cast into this useful geometry for cluster-assembled materials.

\section{Acknowledgements}

Funding within the DFG Cluster of Excellence Unifying Concepts in Catalysis and the Research Unit FOR1282 is gratefully acknowledged.
We thank Tobias Lau for fruitful discussions concerning their XAS experiments, and Volker Blum for FHI-aims technical support.

\end{document}